\documentclass[journal]{IEEEtran}

\usepackage{amsmath}
\usepackage{amsfonts}
\usepackage{graphicx}
\usepackage{hyperref}
\usepackage{booktabs}
\usepackage{microtype}
\usepackage{xcolor}
\usepackage{url}
\usepackage{cite}

\begin{document}

\title{Trajectory Dynamics in Self-Supervised Learning Latent Space for Audio Deepfake Detection}

\author{Tomás Andrade Weber%
  \thanks{T.\ Andrade Weber is with the Barcelona Supercomputing Center (BSC), CASE Department, Plaza Eusebi Güell, 1-3, 08034 Barcelona, Spain (e-mail: tomas.andrade@bsc.es).}%
  \thanks{This work was supported by BSC AI4Science Fellowship. All computations were carried out in Marenostrum5 at the Barcelona Supercomputing Center.}}

\markboth{IEEE Signal Processing Letters}%
{Andrade Weber: Trajectory Dynamics in SSL Latent Space for Audio Deepfake Detection}

\maketitle

\begin{abstract}
Human speech production is constrained by physiology, giving rise to
characteristic temporal structure on acoustic signals. We hypothesise
that these constraints manifest as structured trajectory dynamics in
the latent space of Self-Supervised Learning (SSL) models, and that
synthetic speech violates them detectably. To test this hypothesis, we
train a causal Long Short-Term Memory (LSTM) next-frame predictor on
bonafide speech only (Stage~1), using the deepfake-specialised SSL
backbone Wav2Vec2-Large-AntiDeepfake, and compare against a static
global-average-pooling baseline using identical features, thus
isolating the contribution of temporal modelling. A supervised
Stage~2, which trains a Multi-Layer Perceptron on the frozen LSTM
internal states using labelled data, is included to characterise the
role of spoof supervision. Our system achieves competitive or
state-of-the-art performance across six benchmarks: ASVspoof
2019/2021, Codecfake, In-the-Wild, MLAAD-EN, and
Deepfake-Eval-2024, including best published EER on ASVspoof~2021
(0.75\%) and, notably, Stage~1 trained on bonafide speech only
surpasses the published supervised baseline from the same backbone on
DE2024 (30.35\%). On near-domain benchmarks, static and dynamic
approaches perform comparably. On harder cross-corpus benchmarks with
diverse synthesis methods, trajectory dynamics provide substantial
gains, confirming that temporal physiological constraints carry
detection signal beyond utterance-level statistics.
\end{abstract}

\begin{IEEEkeywords}
audio deepfake detection, self-supervised learning, trajectory dynamics,
one-class learning, physiological constraints
\end{IEEEkeywords}

\IEEEpeerreviewmaketitle

\section{Introduction}
\label{sec:intro}

\IEEEPARstart{T}{he} rapid proliferation of systems capable of
producing realistic synthetic voices, whether Text-to-Speech (TTS) or
Voice Conversion (VC), carries a sharp societal risk. A community has
emerged to provide reliable mechanisms to detect such deepfakes,
producing common benchmarks such as the pioneering ASVspoof
series~\cite{wang2020asvspoof2019, yamagishi2021asvspoof2021} and
increasingly challenging real-world
evaluations~\cite{muller2022inthewild, chandra2025de2024}.

State-of-the-art approaches such as AntiDeepfake~\cite{ge2025posttraining},
SLIM~\cite{SLIM}, QAMO~\cite{truong2025qamo}, and
BreathNet~\cite{ye2026breathnet} rely on SSL feature extractors such
as wav2vec~2.0~\cite{NEURIPS2020_92d1e1eb} or
XLS-R~\cite{babu22_interspeech} as frontends. These are strong on
matched benchmarks, but generalisation to diverse real-world synthesis
remains an open problem: performance degrades severely on hard
cross-corpus benchmarks such as MLAAD~\cite{muller2024mlaad} and
Deepfake-Eval-2024 (DE2024)~\cite{chandra2025de2024}, where the
backbone representations alone are insufficient to separate bonafide
from synthetic speech.

Most SSL-based systems aggregate frame-level embeddings via global
average pooling or attention-weighted summarisation, discarding
temporal order entirely. A smaller body of work exploits temporal
structure directly: FGFM~\cite{truong2026fgfm} selects locally
suspicious frames via attention voting, TRACE~\cite{trace2026} measures
first-order embedding velocity at splice boundaries for partial
deepfake detection, and BreathNet~\cite{ye2026breathnet} guides
training with explicit breath-frame annotations. However, these
approaches target localised artifacts or require spoof supervision.
None models the global physiological plausibility of the full
utterance trajectory.

Prior work has shown that SSL latent spaces organise along patterns
anchored in breath and articulation across
languages~\cite{10.1145/3816077}, suggesting that physiological
production constraints leave a detectable imprint in SSL geometry. We
hypothesise that these constraints manifest as structured trajectory
dynamics, and that synthetic speech, generated without physiological
grounding, violates them detectably. To test this hypothesis, we train
a causal LSTM next-frame predictor on
AntiDeepfake~\cite{ge2025posttraining} features using the
bonafide-only subset of ASVspoof~2019 (Stage~1). A second stage
(Stage~2) reads the frozen LSTM hidden states and trains a supervised
MLP classifier on bonafide and spoof data from ASVspoof~2019. For
each benchmark, we provide a controlled comparison against a static
baseline trained on identical features, isolating the contribution of
temporal modelling.

Our system outperforms SLIM, QAMO, and BreathNet on all shared
benchmarks without backbone fine-tuning, with trajectory dynamics
providing increasing advantage over the static baseline as benchmark
difficulty grows --- from near parity on ASVspoof to large gains on
MLAAD-EN and DE2024. UMAP~\cite{McInnes2018} visualisation of the SSL
embedding space provides mechanistic support for this finding.

\section{Related Work}
\label{sec:related}

\subsection{SSL-based Deepfake Detection}

Self-supervised speech representations have become the dominant
frontend for audio deepfake detection, providing rich phonetic and
speaker embeddings without task-specific supervision. Simple
classifiers on frozen wav2vec~2.0, HuBERT, or XLS-R embeddings
already deliver strong detection, confirming that these models encode
rich forensic information~\cite{serrano2025layer}. Building on this,
Ge et al.\ introduce AntiDeepfake~\cite{ge2025posttraining}, a series
of post-trained SSL models developed on over 56,000 hours of genuine
and 18,000 hours of synthetic speech. The resulting backbone exhibits
strong zero-shot generalisation and serves as the frontend in our
system.

\subsection{One-Class and Anomaly-Based Detection}

A recurring theme in deepfake detection is learning a compact
representation of bonafide speech and flagging deviations as synthetic.
OC-Softmax~\cite{zhang2021ocsoftmax} applies one-class learning with
an angular margin loss, and SAMO~\cite{ding2023samo} extends this with
speaker-attractor multi-centre objectives. More recently,
QAMO~\cite{truong2025qamo} introduces quality-aware multi-centroid
learning, splitting bonafide into high- and low-quality subsets based
on MOS scores. All of these methods define anomaly scores from static
utterance-level distances to learned centroids. Our approach shares the
one-class philosophy of Stage~1 but differs fundamentally in the
anomaly signal: rather than measuring distance to a static centroid, we
model the temporal evolution of the bonafide trajectory and score
utterances by their next-frame prediction error --- capturing dynamics
that static pooling discards. We note that both QAMO and our system
rely on a backbone post-trained on synthetic
speech~\cite{ge2025posttraining}; the one-class claim applies strictly
to the downstream trajectory model.

\subsection{Temporal Modelling}

The temporal dimension of speech has received growing attention as a
complementary signal to spectral features.
FGFM~\cite{truong2026fgfm} proposes fine-grained frame selection via
multi-head voting to identify locally suspicious frames, treating
artifacts as spatially localised. BreathNet~\cite{ye2026breathnet}
explicitly detects breath frames and uses those locations to modulate
XLS-R features during training, finding that the backbone implicitly
encodes breath-related physiological cues even at inference without the
breath mask --- consistent with our hypothesis that SSL representations
carry physiological structure. ProSDD~\cite{mahapatra2026prosdd} learns
speaker-conditioned prosodic variation from real speech as an auxiliary
objective. TRACE~\cite{trace2026} exploits first-order embedding
velocity for training-free detection of partial deepfakes at splice
boundaries. Our work differs from all of these: we target full
deepfakes where no splice boundaries exist, and model the global
physiological plausibility of the entire utterance trajectory rather
than identifying localised artifacts or measuring frame-to-frame
velocity magnitude.

\subsection{Cross-Corpus Generalisation}

Generalisation to out-of-domain synthesis methods remains a central
challenge~\cite{muller2022inthewild}. SLIM~\cite{SLIM}
addresses this through style-linguistics mismatch detection, achieving
competitive EER across multiple benchmarks. Yang et
al.~\cite{yang2025riskaware} improve upon SLIM through risk-aware style
alignment and structural empirical risk minimisation. Recent work has
further investigated backbone selection~\cite{serrano2025layer} and
latent space augmentation~\cite{huang2025latent} as strategies for OOD
robustness. Our approach is complementary: rather than adapting the
representation space, we model temporal trajectory dynamics within a
fixed backbone, achieving lower EER than all of these systems on shared
benchmarks without any fine-tuning.

\section{Method}
\label{sec:method}

\subsection{AntiDeepfake Features and Static Baseline}

In all experiments, we extract speech features using
Wav2Vec2-Large-AntiDeepfake~\cite{ge2025posttraining}, which provides
1024-dimensional frame-level representations at 50 frames per second.
These features are post-trained to discriminate bonafide from synthetic
speech, providing a strong representational foundation even before
temporal modelling.

To quantify the contribution of trajectory dynamics independently of
backbone quality, we construct a static baseline using the same
features. For each dataset, we compute the centroid of all bonafide
training utterance vectors and score each evaluation utterance by the
L2 distance between its global average pooled (GAP) embedding and this
centroid. This directly mirrors the global average pooling readout used
in the original AntiDeepfake evaluation~\cite{ge2025posttraining},
discarding all temporal order information.

\subsection{Stage 1: Causal LSTM Trajectory Predictor}

We model trajectory dynamics using a 2-layer Long Short-Term Memory
(LSTM) network with hidden dimension 512. Given an ordered sequence of
$T$ frame embeddings $\mathbf{h}_1, \ldots, \mathbf{h}_T \in
\mathbb{R}^{1024}$ extracted from a single utterance, the LSTM
causally predicts each next frame from its predecessors. The model is
trained exclusively on the 2{,}580 bonafide utterances of the
ASVspoof~2019 LA training set, minimising the mean squared error
between predicted and true next frames. The resulting per-utterance
anomaly score is:
\begin{equation}
  s(\mathbf{x}) = \frac{1}{T-1} \sum_{t=1}^{T-1}
          \left\| \hat{\mathbf{h}}_t - \mathbf{h}_{t+1} \right\|^2
  \label{eq:score}
\end{equation}
\noindent where $\hat{\mathbf{h}}_t =
f_\theta(\mathbf{h}_1,\ldots,\mathbf{h}_t)$ is the causal LSTM
prediction at step $t$. Higher scores indicate trajectories that
deviate from the bonafide dynamics learned during training.

\subsection{Stage 2: Supervised MLP}

As an optional second stage, we use the frozen Stage~1 LSTM as a
feature extractor. The mean-pooled hidden state across all frames
yields a fixed-length 512-dimensional utterance representation. We
extract these representations for all 25{,}380 utterances (2{,}580
bonafide and 22{,}800 spoof) of the ASVspoof~2019 LA training set, and
train a three-layer MLP with dimensions
$[512 \rightarrow 256 \rightarrow 128 \rightarrow 1]$ and dropout
$p=0.3$ to output a spoof probability. Stage~2 introduces spoof
supervision, in contrast to the purely one-class Stage~1.

\section{Experiments}
\label{sec:experiments}

\subsection{Datasets}

We employ ASVspoof~2019 LA train~\cite{wang2020asvspoof2019} as our
training set, using only bonafide speech (2{,}580 utterances) for
Stage~1, while Stage~2 additionally uses the 22{,}800 spoof utterances.
We evaluate on six benchmarks summarised in Table~\ref{tab:datasets}.
ASVspoof~2019 LA eval and ASVspoof~2021 DF
eval~\cite{yamagishi2021asvspoof2021} provide controlled evaluation on
studio-quality TTS and VC attacks. In-the-Wild~\cite{muller2022inthewild}
contains spontaneous celebrity speech with unknown synthesis methods.
Codecfake~\cite{10830534} targets LLM-based neural codec
synthesis. MLAAD-EN~\cite{muller2024mlaad} is the English subset
(en\_US + en\_UK) of MLAAD v9 with M-AILABS as bonafide source.
Deepfake-Eval-2024 (DE2024)~\cite{chandra2025de2024} is a challenging
real-world benchmark of social media deepfakes; DE2024 audio is split
into non-overlapping 10-second segments following the NII evaluation
protocol~\cite{ge2025posttraining}. Notably, ASVspoof~2019/2021,
Codecfake, and MLAAD spoof audio are all included in the backbone
post-training data~\cite{ge2025posttraining}; only In-the-Wild and
DE2024 are fully out-of-distribution for the backbone. For MLAAD, the
backbone saw only the synthetic audio; the bonafide M-AILABS speech
used as our reference was not in its training set, making the bonafide
trajectory model genuinely zero-shot. Prior to feature extraction,
utterances are preprocessed to remove long silences exceeding 500\,ms,
replacing them with 200\,ms natural pauses, to eliminate recording
padding artifacts present in the
ASVspoof~2019 corpus~\cite{wang2020asvspoof2019} that can act as
spurious detection shortcuts~\cite{muller2021silence}.

\begin{table}[t]
\centering
\caption{Evaluation datasets. $^\dagger$Training set for Stage~1
         (bonafide only) and Stage~2 (bonafide + spoof).
         $^\ddagger$MLAAD-EN v9 (en\_US + en\_UK); bonafide from
         M-AILABS. SLIM uses MLAAD-EN v3 (37,998 utterances).
         $^\S$ASVspoof 2019/2021, Codecfake, and MLAAD (spoof only)
         are in the backbone training data; only ITW and DE2024 are
         fully out-of-distribution.}
\label{tab:datasets}
\setlength{\tabcolsep}{4pt}
\begin{tabular}{lrrr}
\toprule
\textbf{Dataset} & \textbf{Bonafide} & \textbf{Spoof} & \textbf{Hours} \\
\midrule
ASVspoof 2019 train$^\dagger$ & 2{,}580  & 22{,}800  & $\sim$8   \\
\midrule
ASVspoof 2019 eval  & 7{,}355   & 63{,}882   & $\sim$28  \\
ASVspoof 2021 eval  & 22{,}617  & 589{,}212  & $\sim$670 \\
Codecfake$^\S$      & 20{,}000  & 140{,}000  & $\sim$160 \\
In-the-Wild         & 11{,}816  & 19{,}963   & $\sim$38  \\
MLAAD-EN$^\ddagger$ & 69{,}855  & 116{,}000  & $\sim$220 \\
DE2024              & 4{,}712   & 2{,}543    & $\sim$21  \\
\bottomrule
\end{tabular}
\end{table}

\subsection{Implementation Details}

We train the Stage~1 LSTM for 500 epochs using AdamW with cosine
annealing and warm restarts ($T_0 = 50$, $T_\text{mult} = 2$),
initial learning rate $10^{-3}$, dropout $p=0.1$, and maximum sequence
length 500 frames. Stage~2 MLP is trained for 100 epochs with
dropout $p=0.3$. All experiments use a single H100 GPU. Performance
is reported using Equal Error Rate (EER\,\%), consistent with
the ASVspoof challenge protocol.
Code and configurations are available on Zenodo \footnote{\url{https://doi.org/10.5281/zenodo.21879214}}.

To estimate variance, we train five independent Stage~1 models using 5 random seeds, and for each Stage~1 model we
train five independent Stage~2 MLP classifiers, yielding 25 runs per benchmark. The static baseline is deterministic and requires no repeated runs. Reported mean and standard deviation are computed across all 25 runs, capturing both LSTM initialisation variance and MLP training variance. Stage~1 standard deviation is computed from the
same 25 runs by averaging Stage~2 scores within each Stage~1 seed before computing statistics across seeds.

\section{Results}
\label{sec:results}
Table~\ref{tab:results} compares our system against published baselines
across six benchmarks. On near-domain benchmarks (ASVspoof~2019/2021,
Codecfake), Stage~2 performs comparably to the static baseline and
significantly better than Stage~1; Stage~2 outperforms all published
competitors on these benchmarks. On out-of-domain benchmarks (MLAAD-EN,
DE2024), Stage~1 outperforms the static baseline by a large margin and
outperforms Stage~2. On DE2024, Stage~1 surpasses the NII-GAP baseline
despite using no spoof supervision, while Stage~2 converges to
performance similar to other supervised systems trained on the same
data.

\subsection{The Difficulty Gradient}

On ASVspoof~2019/2021 and Codecfake, the gap between the static
baseline and Stage~2 is small ($<$0.5\% EER), indicating that the
backbone representations largely suffice to separate bonafide from
synthetic speech. The gap opens on In-the-Wild ($-$0.75\% EER) where
Stage~2 retains an advantage over Stage~1. On the hardest
out-of-domain benchmarks, MLAAD-EN and DE2024, Stage~1 provides the
best result. The advantage of Stage~1 over the static baseline grows
from 17.2\% on MLAAD-EN to 22.2\% on DE2024, confirming that
trajectory dynamics become increasingly important as benchmark
difficulty grows.

\begin{figure*}[t!]
  \centering
  \includegraphics[width=\linewidth]{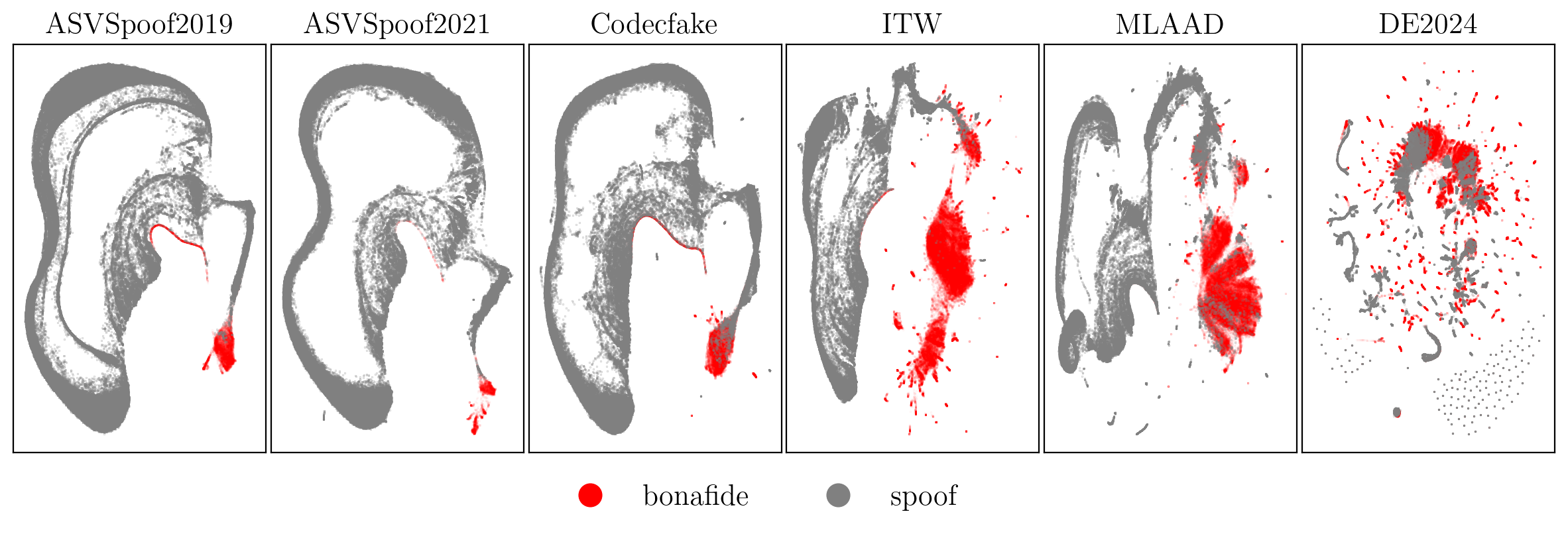}
  \caption{UMAP visualisation of AntiDeepfake embedding space (red:
    bonafide, grey: spoof) across six benchmarks in increasing order
    of difficulty (left to right). Bonafide/spoof overlap grows
    progressively from tight separation (ASVspoof 2019/2021, Codecfake)
    through partial overlap (In-the-Wild, MLAAD-EN) to near-complete
    mixing (DE2024), explaining the corresponding degradation of the
    static baseline and the growing advantage of trajectory dynamics.}
  \label{fig:umap}
\end{figure*}

This progression is directly visible in the UMAP projections of
Fig.~\ref{fig:umap}, computed from $\sim$175,000 randomly sampled
frames per dataset using UMAP~\cite{McInnes2018} with parameters
$n_\text{neighbors}{=}15$, $\text{min\_dist}{=}0.1$, cosine metric.
ASVspoof~2019/2021 and Codecfake show a tight, well-separated bonafide
cluster, consistent with their low static EER. In-the-Wild shows a
larger, more diffuse bonafide distribution with partial overlap.
MLAAD-EN exhibits heavy overlap, and DE2024 shows near-complete mixing
of bonafide and spoof representations --- explaining why the static
baseline fails at 22.9\% and 52.5\% EER respectively. It is precisely
in these harder regimes where trajectory dynamics provide the critical
discriminative signal. Note also that Stage~2 variance is larger on
harder benchmarks (DE2024: $\pm$2.62\%, MLAAD-EN: $\pm$1.22\%),
reflecting sensitivity to the Stage~1 bonafide manifold estimate when
bonafide/spoof overlap is substantial --- consistent with the
difficulty gradient observed in Fig.~\ref{fig:umap}.

\subsection{One-Class vs Supervised Generalisation}

The reversal between Stage~1 and Stage~2 on harder benchmarks is
informative. Stage~2 supervision is derived from ASVspoof~2019 attack
types (A01--A06), which are not representative of the 140+ TTS systems
in MLAAD-EN or the real-world deepfakes in DE2024. The one-class
Stage~1, trained only on bonafide speech, is not affected by this
mismatch and generalises more robustly to unseen synthesis conditions.
Notably, on both MLAAD-EN and DE2024, Stage~2 converges to performance
comparable to published supervised baselines (SLIM: 10.7\% and
NII-GAP: 33.36\%), suggesting that the ASVspoof~2019 supervision
signal, rather than the architecture, is the bottleneck for
cross-domain generalisation.

\begin{table*}[t]
\centering
\caption{EER (\%) comparison across benchmarks. Lower is better.
  \emph{Static} uses the same AntiDeepfake backbone with global average
  pooling and is deterministic. Stage~1 and Stage~2 report mean\,$\pm$\,std
  over 25 independent runs (5 Stage~1 seeds $\times$ 5 Stage~2 MLP runs).
  $^\dagger$NII-GAP: same backbone with GAP+FC and full spoof
  supervision~\cite{ge2025posttraining}.
  $^\ddagger$MLAAD-EN v9 (185K utterances); SLIM uses v3 (38K).
  $^\S$ASVspoof 2019/2021, Codecfake, and MLAAD (spoof only) are in
  the backbone training data; only ITW and DE2024 are fully
  out-of-distribution.
  $^*$NII-GAP DE2024 uses whole-file inference.}
\label{tab:results}
\setlength{\tabcolsep}{3.5pt}
\small
\begin{tabular}{lcccccccc}
\toprule
\textbf{Benchmark}
  & \textbf{Static}
  & \textbf{Stage 1}
  & \textbf{Stage 2}
  & \textbf{BreathNet}~\cite{ye2026breathnet}
  & \textbf{QAMO}~\cite{truong2025qamo}
  & \textbf{SLIM}~\cite{SLIM}
  & \textbf{NII-GAP}$^\dagger$~\cite{ge2025posttraining} \\
  & \multicolumn{3}{c}{\textit{(ours)}} & & & & \\
\midrule
ASVspoof 2019
  & 1.51
  & $2.57 \pm 0.12$
  & $\mathbf{1.11} \pm 0.10$
  & 0.23  & ---  & 0.30  & ---       \\
ASVspoof 2021
  & 0.98
  & $1.88 \pm 0.05$
  & $\mathbf{0.75} \pm 0.10$
  & 1.87  & 1.54 & 3.60  & ---       \\
Codecfake$^\S$
  & 3.21
  & $5.21 \pm 0.09$
  & $\mathbf{2.43} \pm 0.18$
  & ---   & ---  & ---   & ---       \\
In-the-Wild
  & 4.03
  & $4.84 \pm 0.04$
  & $\mathbf{3.28} \pm 0.60$
  & 4.70  & 5.09 & 12.5  & 1.91      \\
MLAAD-EN$^\ddagger$
  & 22.86
  & $\mathbf{5.71} \pm 0.03$
  & $10.02 \pm 1.22$
  & ---   & ---  & 10.7  & ---       \\
DE2024
  & 52.52
  & $\mathbf{30.35} \pm 0.15$
  & $35.41 \pm 2.62$
  & ---   & ---  & ---   & 33.36$^*$ \\
\bottomrule
\end{tabular}
\end{table*}

\section{Conclusion}
\label{sec:conclusion}

Our results demonstrate that trajectory dynamics in AntiDeepfake SSL
space provide a reliable one-class detection signal that static pooling
cannot capture, with the advantage scaling with benchmark difficulty.
Moreover, the reversal from Stage~1~$>$~Stage~2 on hard
out-of-domain benchmarks indicates that one-class bonafide modelling
generalises better than supervised spoof classification to unseen
synthesis methods. We leave for future work the comparison of other
sequence-based algorithms such as bidirectional LSTM or Transformers,
fine-tuning with DE2024 in-domain training, and multilingual
generalisation.

\bibliographystyle{IEEEtran}
\bibliography{refs}

\end{document}